# Incorporating nuclear vibrational energies into the "atom in molecules" analysis: An analytical study

Masumeh Gharabaghi[1] and Shant Shahbazian[2,*]

[1]*Faculty of Chemical and Petroleum Sciences, Shahid Beheshti University, G. C., Evin, Tehran, Iran, 19839, P.O. Box 19395-4716.*

[2] *Department of Physics, Shahid Beheshti University, G. C., Evin, Tehran, Iran, 19839, P.O. Box 19395-4716.*

E-mails:
(Shant Shahbazian)
sh_shahbazian@sbu.ac.ir
chemist_shant@yahoo.com

[*] Corresponding author

**Abstract**

The quantum theory of atoms in molecules (QTAIM) is based on the clamped nucleus paradigm and solely working with the electronic wavefunctions, so does not include nuclear vibrations in the AIM analysis. On the other hand, the recently extended version of the QTAIM, called the multi-component QTAIM (MC-QTAIM), incorporates both electrons and quantum nuclei, i.e. those nuclei treated as quantum waves instead of clamped point charges, into the AIM analysis using non-adiabatic wavefunctions. Thus, the MC-QTAIM is the natural framework to incorporate the role of nuclear vibrations into the AIM analysis. In this study, within the context of the MC-QTAIM, the formalism of including nuclear vibrational energy in the atomic basin energy is developed in detail and its contribution is derived analytically using the recently proposed non-adiabatic Hartree product nuclear wavefunction. It is demonstrated that within the context of this wavefunction the quantum nuclei may be conceived pseudo-adiabatically as quantum oscillators and both isotropic harmonic and anisotropic anharmonic oscillator models are used to compute the zero-point nuclear vibrational energy contribution to the basin energies explicitly. Inspired by the results gained within the context of the MC-QTAIM analysis, a heuristic approach is proposed within the context of the QTAIM to include nuclear vibrational energy in the basin energy from the vibrational wavefunction derived adiabatically. The explicit calculation of the basin contribution of the zero-point vibrational energy using uncoupled harmonic oscillator model leads to results consistent with those derived from the MC-QTAIM.

# I. Introduction

The multi-component quantum theory of atoms in molecules (MC-QTAIM) is the extension of the QTAIM,[1-3] aiming to widen the domain of the "atoms in molecules" (AIM) analysis.[4-18] This widening includes the AIM analysis of the usual molecular species beyond the clamped nuclei paradigm and various exotic molecules, i.e. positronic and muonic species, as well. The AIM analysis of the exotic species is a completely novel dimension, which is only materialized within the context of the MC-QTAIM since in these species not only electrons but also exotic particles, i.e. positrons or positively charged muons, are treated as quantum waves instead of clamped point charges. In other words, the kinetic energy operators of both electrons and the exotic particle appear in corresponding Schrödinger equation and the resulting wavefunction contains the variables of both types of particles. The QTAIM is inherently a *single-component* theory, inevitably rooted in the clamped nuclei paradigm, only dealing with the electronic wavefunctions and unable to derive the AIM structure of systems containing two or more types of quantum particles. Besides, in usual molecular species one may treat certain light nuclei as quantum waves, e.g. the isotopes of hydrogen, portraying molecule as a two or even *multi-component* quantum system. Evidently, both exotic species and those molecular systems considered beyond the clamped nuclei paradigm have a similar facet, a multi-component wavefunction that cannot be used as an "input" to the QTAIM to unveil the underlying AIM structure. This is where the MC-QTAIM enters the scene; employing the multi-component wavefunctions as "input" and delivering the AIM structure, i.e., both morphology and properties, as "output". Indeed, the last fifteen years witness enormous progress in the development of various ab initio procedures that aim to solve the multi-component Schrödinger equation, treating electrons and nuclei/exotic particle from the *outset* as quantum waves without invoking the clamped nucleus paradigm.[19-29] The resulting ab initio multi-component wavefunctions are the raw material for the MC-QTAIM analysis which deciphers part of the "chemical content" of these wavefunctions. Although just the tip of an iceberg has been explored,[4-18] the handful of systems considered computationally within the context of the MC-QTAIM reveal subtle effects that are absent in the QTAIM analysis. Examples include: I- Disclosing the uneven distribution of positrons in atomic basins and quantifying regional positron affinities,[4-6] II- Disclosing dissimilar electronegativity of the hydrogen isotopes,[7,15] III- Demonstrating the capacity of the positively charged muon to form its own atomic basin,[13,16,17] IV- Disclosing the subtle effects of the isotope substitution on the covalent character of chemical bonds,[18] V- Discovering the novel phenomena of the topological mass-induced structural transitions,[14] all unconceivable within the context of the QTAIM.

Beyond the computational considerations, the MC-QTAIM formalism offers new theoretical possibilities for "analytical" studies of various basin properties, e.g. the energy of AIM, which rarely have been pursued within the context of the QTAIM.[30] This stems from the fact that the electronic and the nuclear contributions in basin properties are distinct and in contrast to the delocalized nature of the electrons, in most cases the nuclei are rather localized quantum particles and usually well-described with simplified localized wavefunctions. It is relatively straightforward to perform analytical studies on the nuclear contributions, apart from the electronic contributions, using localized wavefunctions particularly when the total translational-rotational motions of nuclei are discarded and just the vibrational motions are in focus. As an example, using a Hartree product nuclear wavefunction proposed by Auer and Hammes-Schiffer,[31] the explicit mass-dependence of the nuclear contribution of various basin properties was derived;[9] more recently, it was demonstrated that Hartree product nuclear wavefunction may be interpreted "pseudo-adiabatically" and in this scheme the localized nuclei act as quantum oscillators.[32] Interestingly, in the asymptotic case when the mass of a nucleus is infinite, the nuclear contributions to various basin properties are null and results of the MC-QTAIM analysis are indistinguishable from the QTAIM.[9] This finding, because of its analytical nature, is quite general and points to the fact that the QTAIM emerges as an asymptote of the MC-QTAIM therefore in the large nuclear mass limit, the MC-QTAIM recovers all the known results of the QTAIM.[9] In another analytical study the explicit mass-dependence of the localization/delocalization indices of the quantum nuclei was derived and it was demonstrated that a massive quantum nucleus resides in its own atomic basin with no tendency to be delocalized into neighboring basin.[11] This is also in line with the emerging image of the AIM structure within the context of the QTAIM where each atomic basin is composed of a clamped nucleus and surrounding electronic population within the atomic basin.[1-3] This report is also a continuation of the previous analytical studies that were confined to the two-component systems.[9,11]

Generally, the molecular systems considered within context of the MC-QTAIM can be categorized into two classes: A- Molecules containing quantum nuclei/exotic particles with low-amplitude vibrations around their equilibrium centers where the distribution of each quantum nucleus/exotic particle is effectively confined to a single atomic basin, B- Molecules containing quantum nuclei/exotic particles with large-amplitude vibrations or intra-molecular tunneling, where the distributions of these particles are not localized around a single center in 3D space and penetrate into neighboring atomic basins beyond their own basins. Present study focuses on the class A trying to incorporate nuclear vibrational energies into the atomic basin energies analytically, employing the Hartree product nuclear wavefunction, leaving a thorough analysis of the class B to a future study.

## II. Introducing the basin energies within the context of the MC-QTAIM

### A. Background

The formalism of the MC-QTAIM has been discussed thoroughly elsewhere,[10] and herein, only its basic tenets are reiterated briefly before introducing the basin energies. Let's assume a molecular system composed of $p$ types of quantum particles (type 1 is the set of electrons), where $N_n$ is the number of particles for each type ($n = 1,...,p$), all within the external coulombic field generated by $m$ clamped nuclei. The Schrödinger equation, governing the stationary states of the system, is as the following:

$$\hat{H}\Psi = E\Psi$$

$$\hat{H} = \sum_n^p \hat{T}_n + \hat{V}, \quad \hat{V} = \sum_n^p \hat{V}_{nn} + \sum_n^p \sum_{n'>n}^p \hat{V}_{nn'} + \sum_n^p \hat{V}_{nc} + V_{cc}$$

$$\hat{T}_n = \left(-\frac{\hbar^2}{2m_n}\right)\sum_i^{N_n} \nabla_n^{i2}, \quad \hat{V}_{nn} = \sum_i^{N_n} \sum_{j>i}^{N_n} \frac{Q_n^2}{r_n^{ij}}, \quad \hat{V}_{nn'} = \sum_i^{N_n} \sum_j^{N_{n'}} \frac{Q_n Q_{n'}}{r_{nn'}^{ij}},$$

$$\hat{V}_{nc} = \sum_\alpha^m \sum_i^{N_n} \frac{Q_n Z_\alpha}{r_{n\alpha}^i}, \quad V_{cc} = \sum_\alpha^m \sum_{\beta>\alpha}^m \frac{Z_\alpha Z_\beta}{r_{\alpha\beta}} \quad (1)$$

In this equation $Q$ and $Z$ are the electric charges of the quantum particles and clamped nuclei, respectively, while $m_n$ is the mass of type $n-th$ quantum particles. Also, the inter-particle distances are as the following: $r_n^{ij} = |\vec{r}_n^i - \vec{r}_n^j|$, $r_{nn'}^{ij} = |\vec{r}_n^i - \vec{r}_{n'}^j|$, $r_{n\alpha}^i = |\vec{r}_n^i - \vec{R}_\alpha|$, $r_{\alpha\beta} = |\vec{R}_\alpha - \vec{R}_\beta|$. This Hamiltonian is the basis for introducing the MC-QTAIM theorems in all subsequent discussions.

The boundaries delineating the atomic basins of the molecular system are derived from the following equation:

$$\vec{\nabla}\Gamma(\vec{q}).\vec{n}(\vec{q}) = 0$$

$$\Gamma(\vec{q}) = \sum_{n}^{p}\left(\frac{m_1}{m_n}\right)\rho_n(\vec{q}), \qquad \rho_n(\vec{r}_n^{\,1}) = N_n\int d\tau'_n \Psi^*\Psi \qquad (2)$$

$\rho_n(\vec{r}_n^{\,1})$ stands for the one-particle density while $d\tau'_n$ implies summing over spin variables of all quantum particles and integrating over spatial coordinates of all quantum particles except the coordinate of one quantum particle corresponding to $n-th$ type. $\Gamma(\vec{q})$ is called the Gamma density and equation (2) is the local zero-flux equation while $\vec{n}(\vec{q})$ is the unit vector normal to the surface (the justification behind introducing the Gamma density instead of the one-electron density, used within the context of the QTAIM, into the zero-flux equation has been disclosed previously,[10] and is not reiterated herein). It is important to realize that the construction scheme of the Gamma density is the typical approach used to construct all other densities within the context of the MC-QTAIM; the combination strategy involves separate construction of the one-particle densities then using a "unified" coordinate, $\vec{q}$, to combine them and producing the final combined density.[10] The surfaces satisfying the zero-flux equation and going through the (3, -1) critical points of $\vec{\nabla}\Gamma(\vec{q})$ are the boundaries delineating the atomic basins. Each basin, denoted hereafter as $\Omega$, is the union of a (3, -3) critical point of $\vec{\nabla}\Gamma(\vec{q})$ and its attracting basin.

To each mechanical property, $M$, a property density is attributed, $\rho_M^n(\vec{r}_n^{\,1})$, which is constructed according to the combination strategy and then integrated within the boundaries of the target basin to yield the basin contribution, $\tilde{M}(\Omega)$:

$$\tilde{M}(\Omega) = \int_\Omega d\vec{q}\, \tilde{\rho}_M(\vec{q}), \qquad \tilde{\rho}_M(\vec{q}) = \sum_n^P \rho_M^n(\vec{q}) \qquad (3)$$

The sum of basin contributions yields the total property attributed to the whole molecule: $M_{total} = \sum_\Omega \tilde{M}(\Omega)$. To derive the proper property density the local hypervirial theorem is used:

$$\tilde{\rho}_M(\vec{q}) = \vec{\nabla} \bullet \vec{J}_G(\vec{q})$$

$$\rho_M^n(\vec{r}_n^1) = \mathrm{Re}\left\{N_n \int d\tau_n' \Psi^* \left(\frac{i}{\hbar}\right)\left[\hat{H}, \hat{G}_n^1\right]\Psi\right\}, \qquad \vec{J}_G(\vec{q}) = \sum_{n=1}^P \vec{J}_G^n(\vec{q})$$

$$\vec{J}_G^n(\vec{r}_n^1) = \mathrm{Re}\left\{\left(\frac{N_n \hbar}{2m_n i}\right)\int d\tau_n' \left\{\Psi^* \vec{\nabla}_n^1 \bullet \left(\hat{G}_n^1 \Psi\right) - \left(\hat{G}_n^1 \Psi\right)\bullet\left(\vec{\nabla}_n^1 \Psi^*\right)\right\}\right\} \quad (4)$$

In this equation $i = \sqrt{-1}$, and the symbol $\bullet$ is used to emphasize on the dyadic nature of the product while $\hat{G}_n^1$ stands for the one-particle Hermitian operator for $n-th$ type of the quantum particles, used as the generator of the property $\hat{M}$. Employing the following generator: $\hat{G}_n^1 = \left(\frac{1}{2}\right)\left(\vec{r}_n^1 \cdot \vec{p}_n^1 + \vec{p}_n^1 \cdot \vec{r}_n^1\right)$, the "local" virial theorem results:

$$2\tilde{T}(\vec{q}) = -\tilde{V}^T(\vec{q}) - \left(\frac{\hbar^2}{4m_1}\right)\nabla^2 \Gamma(\vec{q})$$

$$T_n(\vec{r}_n^1) = N_n \int d\tau_n' \Psi^* \left(-\frac{\hbar^2}{2m_n}\right)\nabla_n^{1^2}\Psi, \qquad \tilde{V}^T(\vec{q}) = \tilde{V}^B(\vec{q}) + \tilde{V}^S(\vec{q})$$

$$V_n^B(\vec{r}_n^1) = N_n \int d\tau_n' \Psi^* \left(-\vec{r}_n^1 \cdot \vec{\nabla}_n^1 \hat{V}\right)\Psi, \qquad V_n^S(\vec{r}_n^1) = \vec{\nabla}_n^1 \cdot \left(\vec{r}_n^1 \bullet \vec{\sigma}_n(\vec{r}_n^1)\right) \quad (5)$$

In these equations $\tilde{T}(\vec{q})$, $\tilde{V}^B(\vec{q})$, $\tilde{V}^S(\vec{q})$, $\tilde{V}^T(\vec{q})$ stand for the combined kinetic energy density, the combined basin virial density, the combined surface virial density, and the combined total virial density, respectively, while $\vec{\vec{\sigma}}_n(\vec{r}_n^1)$ is the Schrodinger-Pauli-Epstein stress tensor density of $n-th$ type of the quantum particles defined as follows:

$$\vec{\vec{\sigma}}_n(\vec{r}_n^1) = \left(\frac{N_n \hbar^2}{4m_n}\right) \int d\tau_n' \left\{ \Psi^*(\vec{\nabla} \bullet \vec{\nabla}\Psi) + \Psi(\vec{\nabla} \bullet \vec{\nabla}\Psi^*) - (\vec{\nabla}\Psi^*) \bullet (\vec{\nabla}\Psi) - (\vec{\nabla}\Psi) \bullet (\vec{\nabla}\Psi^*) \right\}.^{10}$$ For

an atomic basin, enclosed by the zero-flux surfaces, the "regional" virial theorem is as follows:

$$2\tilde{T}(\Omega) = -\tilde{V}^T(\Omega) \tag{6}$$

The last term in the local virial theorem disappears from the regional variant because of Gauss's theorem: $\int_\Omega d\vec{q} \ \nabla^2 \Gamma(\vec{q}) = \oint_{\partial\Omega} dS \ \vec{\nabla}\Gamma(\vec{q}).\vec{n}(\vec{q}) = 0$. Taking this background into account, let's at this stage of development introduce the basin energies.

**B. Basin energies**

In order to introduce the basin energies, the two-particle terms of the coulombic potential energy operator are partitioned into the one-particle contributions and the virial operator is introduced at the first step as the "projection" operator to materialize this partitioning:

$$\hat{V} = -\sum_n^p \sum_i^{N_n} \vec{r}_n^i \cdot \vec{\nabla}_n^i \hat{V} - \sum_\alpha^m \vec{R}_\alpha \cdot \vec{\nabla}_\alpha \hat{V} \tag{7}$$

Employing this operator, the Hamiltonian proposed in equation (1) is transformed as the following:

$$\hat{H} = \sum_n^p \hat{H}_n - \sum_\alpha^m \vec{R}_\alpha \cdot \vec{\nabla}_\alpha \hat{V}$$

$$\hat{H}_n = \sum_i^{N_n} \hat{H}_n^i = \sum_i^{N_n} \left\{ \left( \frac{-\hbar^2}{2m_n} \right) \nabla_n^{i2} - \vec{r}_n^i \cdot \vec{\nabla}_n^i \hat{V} \right\} \qquad (8)$$

According to the Hellmann-Feynman theorem for a molecular system, at the mechanical equilibrium the forces operative on clamped nuclei are null, $\vec{F}_\alpha = -\vec{\nabla}_\alpha \hat{V} = 0$, and the second term in the Hamiltonian disappears: $\hat{H} = \sum_n^p \hat{H}_n$. Now, the energy of the system is partitioned into distinct contributions emerging from each type of the quantum particles:

$$E = \sum_n^p E_n, \qquad E_n = \int d\vec{r}_n^i E_n(\vec{r}_n^i), \qquad E_n(\vec{r}_n^i) = N_n \int d\tau_n' \Psi^* \hat{H}_n^i \Psi \qquad (9)$$

The energy density appeared in the last expression is particularly proper for introducing the energy density within the context of the MC-QTAIM. Accordingly, one may rewrite equations (9) as the following:

$$E = \int d\vec{q}\, \tilde{E}(\vec{q}), \qquad \tilde{E}(\vec{q}) = \sum_n^p E_n(\vec{q}), \qquad E_n(\vec{q}) = T_n(\vec{q}) + V_n^B(\vec{q}) \qquad (10)$$

Although these equations are properly extended, the basin virial is origin-dependent and this is an undesired feature for a consistent definition of basin energies.[9] The clue for a proper modification comes from the local or the regional virial theorems where the surface virial term also appears in the equations, which is absent in equations governing the total molecular system. Accordingly, equations (10) are modified as the following:

$$\tilde{E}(\vec{q}) = \tilde{T}(\vec{q}) + \tilde{V}^T(\vec{q}) \qquad (11)$$

Based on this energy density as well as the regional virial theorem, the basin energy is introduced:

$$\tilde{E}(\Omega) = \tilde{T}(\Omega) + \tilde{V}^T(\Omega) = -\tilde{T}(\Omega) = \left(\frac{1}{2}\right)\tilde{V}^T(\Omega) \tag{12}$$

In the next section employing the developed formalism, the contribution of the nuclear vibrational energies of the quantum nuclei in the basin energies is considered in detail.

## III. Analytical derivation of nuclear vibrational energy contribution to the basin energy

### A. The simplified formalism for the localized quantum nuclei

Equation (12) yields the total energy of an atomic basin where all the quantum particles contribute to the energy while the contribution of each type of the quantum particles is derived as follows:

$$E_n(\Omega) = \int_\Omega d\vec{r}_n^1 \, E_n(\vec{r}_n^1)$$

$$E_n(\vec{r}_n^1) = T_n(\vec{r}_n^1) + V_n^T(\vec{r}_n^1), \quad V_n^T(\vec{r}_n^1) = V_n^B(\vec{r}_n^1) + V_n^S(\vec{r}_n^1) \tag{13}$$

If the ground state wavefunction is used to derive the energy density, then $E_n(\Omega)$, $n>1$, maybe interpreted as the contribution of the vibrational energies of the $n-th$ type of the quantum nuclei in the atomic basin $\Omega$. Let's now assume that one is faced with a molecule from the class A and each quantum nucleus is effectively localized into its own atomic basin, i.e., $\int_{\Omega_n} d\vec{r}_n^1 \, \rho_n(\vec{r}_n^1) \approx 1$ and $\int_{\Omega_{m \neq n}} d\vec{r}_n^1 \, \rho_n(\vec{r}_n^1) \approx 0$, $n>1$.[15] In this case, the quantum nuclei are practically distinguishable and each quantum nucleus could be conceived as a single type, i.e., $N_n = 1$, $n>1$ (since there are $p$ number of the quantum nuclei thus the superscript "1" is redundant, and dropped from all subsequent equations). Now, the interpretation of $E_n(\Omega)$ is even more straightforward since it is the vibrational energy of the $n-th$ quantum nucleus and because of the assumed localization, it is null for all basins except the basin $\Omega_n$ where the nucleus is confined within. Let's now consider the consequences of this localization on the formalism developed in the previous section.

The local virial theorem is the combination of $n-th$ local virial theorems, one for electrons, and $n-1$ for each quantum nucleus:

$$2T_1(\vec{r}_1^1) = -V_1^T(\vec{r}_1^1) - \left(\frac{\hbar^2}{4m_1}\right)\nabla^2 \rho_1(\vec{r}_1^1)$$

$$2T_n(\vec{r}_n) = -V_n^T(\vec{r}_n) - \left(\frac{\hbar^2}{4m_n}\right)\nabla^2 \rho_n(\vec{r}_n), \quad n>1 \tag{14}$$

The logic behind the combination strategy is the fact that in general, after integration within an atomic basin, the last terms do not disappear from the equations and a separate regional virial theorem does not emerge for each type of the quantum particles.[10] However, because of the assumed localization of each quantum nucleus within its own atomic basin, one may claim: $\int_{\Omega_n} d\vec{r}_n \, \nabla^2 \rho_n(\vec{r}_n) \approx 0, \, n>1$. This means that now two separate regional theorems maybe derived for the systems of the class A:

$$2T_n(\Omega_n) = -V_n^B(\Omega_n), \quad n>1,$$

$$2T_1(\Omega_n) = -V_1^T(\Omega_n) - \left(\frac{\hbar^2}{4m_1}\right) \oint_{\partial\Omega_n} dS \, \vec{\nabla}\rho_1(\vec{r}_1^1).\vec{n}(\vec{r}_1^1) \tag{15}$$

In the first equation, the total virial has been transformed into the basin virial since the nuclear stress tensor densities are practically null at boundaries of the basins, $\vec{\vec{\sigma}}_n(\partial\Omega_n) \approx 0$, and consequently: $V_n^S(\Omega_n) = \int_{\Omega_n} d\vec{r}_n \, \vec{\nabla}_n.(\vec{r}_n \bullet \vec{\vec{\sigma}}_n(\vec{r}_n)) = \oint_{\partial\Omega_n} dS(\vec{r}_n \bullet \vec{\vec{\sigma}}_n(\vec{r}_n)).\vec{n} \approx 0$. Also, the boundaries of the atomic basins are derived from the local zero-flux equation of the one-particle density of electrons: $\vec{\nabla}\rho_1(\vec{r}_1^1).\vec{n}(\vec{r}_1^1) = 0$, which replaces the more general zero-flux equation of the Gamma density given in equation (2), and the second equation reduces to: $2T_1(\Omega_n) = -V_1^T(\Omega_n)$. This reasoning justifies the neglect of the quantum particles other than electrons in the zero-flux equation used in the QTAIM to delineate the atomic boundaries; the QTAIM is applicable when the vibrational amplitudes of nuclei are localized enough to neglect their role in the delineation of the basin boundaries. Now, combining equations (12), (13) and (15) one arrives at:

$$\tilde{E}(\Omega_n) = E_1(\Omega_n) + E_n(\Omega_n),$$

$$E_1(\Omega_n) = -T_1(\Omega_n) = \left(\frac{1}{2}\right)V_1^T(\Omega_n)$$

$$E_n(\Omega_n) = -T_n(\Omega_n) = \left(\frac{1}{2}\right)V_n^B(\Omega_n),\ n > 1 \qquad (16)$$

The message delivered by these equations are transparent; the energy of an atomic basin containing a quantum nucleus is the sum of the electronic and the nuclear energies while the latter is the vibrational energy of the nucleus. At this stage of development, the simplified formalism is complete and one may introduce an explicit model wavefunction for the nuclear vibrations and considers the consequences of equation (16).

**B. The isotropic harmonic oscillator model of the quantum nuclei**

In order to calculate the contribution of the nuclear vibrations to the basin energies analytically, a simplified but reliable enough wavefunction must be employed. Accordingly, the Hartree product nuclear wavefunction proposed by Auer and S. Hammes-Schiffer,[31] is used in present study for all subsequent analytical calculations:

$$\Phi = \psi_1\left((\vec{r}_1^{\,1}, \sigma_1^1), \ldots, (\vec{r}_1^{\,N_1}, \sigma_1^{N_1})\right) \prod_{n=2}^{p} \psi_n(\vec{r}_n) \sigma_n, \quad \sigma_1^i = \alpha \text{ or } \beta, \quad \sigma_n = \alpha \quad (17)$$

The nuclear part of the wavefunction is in its highest spin-state (all $\alpha$ spin states maybe changed to $\beta$ states simultaneously without affecting the subsequent discussions) while $\psi_1$ has a quite general nature and in principle, may contain the full electron-electron correlation. The logic behind proposing this wavefunction is the fact that the nuclear wavefunction does not need to be described by a Slater determinant since the nuclear exchange and correlation effects are effectively null because of the assumed localization, as has also been demonstrated computaionally.[31] The electron-nucleus correlation has also been neglected in this wavefunction so this affects the numerical accuracy of the wavefunction if it is used for ab initio quantum mechanical calculations,[33] however, for the present purpose this wavefunction seems to be proper. It is important to emphasize that while this product wavefunction may seem adiabatic-like in the first glance, since in practice the nuclear and electronic wavefunctions are determined through the variational principle "simultaneously",[31] they are non-adiabatic and numerically distinct from their adiabatic counterparts. In order to proceed, $\psi_n(\vec{r}_n)$ must be specified and to do so, a model for nuclear vibrations must be constructed.

In a recent study,[32] it has been proposed that the Hartree product nuclear wavefunction implies that each quantum nucleus is conceived as a quantum oscillator in a hypothetical external field, $V_n^{ext}$, and the following Schrödinger equation governs the oscillator:

$$\left(\frac{-\hbar^2}{2m_n}\right) \nabla_n^2 \psi_n(\vec{r}_n) + V_n^{ext}(\vec{r}_n) \psi_n(\vec{r}_n) = \varepsilon_n \psi_n(\vec{r}_n) \qquad (18)$$

Generally, $V_n^{ext}$ is not *a priori* known and one may "design" an external field based on available non-adiabatic ab initio results. First, Let's consider the simplest conceivable model potential namely, the isotropic harmonic oscillator, $\hat{V}_n^{ext.} = \left(\frac{2\alpha_n^2\hbar^2}{m_n}\right)\left|\vec{r}_n - \vec{R}_{n,c}\right|^2$, where $\alpha_n$ and $\vec{R}_{n,c}$ are the parameters to be determined for each oscillator; the optimized values of the parameters are found using the wavefunction in equation (17) as a trial wavefunction in the variational principle.[31] The ground state wavefunction of the isotropic harmonic oscillator model is: $\psi_n(\vec{r}_n) = g_s(\alpha_n) = \left(\frac{2\alpha_n}{\pi}\right)^{\frac{3}{4}} \exp\left(-\alpha_n\left|\vec{r}_n - \vec{R}_{n,c}\right|^2\right)$, and the product of the s-type Gaussian functions is used as the explicit form of the nuclear wavefunction as is done also in the ab initio calculations.[13-18]

In order to compute the nuclear contribution to the basin energies, according to equations (16), the basin kinetic or virial energies must be computed using the product of the s-type Gaussian functions. After some mathematical manipulation, one arrives at:

$$E_n(\Omega_n) = \frac{-3\alpha_n\hbar^2}{2m_n} \qquad (19)$$

This simple equation makes it possible to consider the effect of the variations of $m_n$ and $\alpha_n$ on the nuclear contribution of the basin energies. Since for an isotropic harmonic oscillator: $\alpha_n = \left(\frac{\sqrt{k_n}}{2\hbar}\right)\sqrt{m_n}$ ($k_n$ stand for the force constant), thus: $\lim_{m_n \to \infty} E_n(\Omega_n) \to 0$. In this limit the clamped nucleus model is materialized and the Gaussian functions practically reduce to the Dirac delta functions and the MC-QTAIM formalism reduces to that of the QTAIM wherein, only electrons contribute to the basin properties. Also, since both $m_n$ and $\alpha_n$ are distinct for each isotope of an atom, the nuclear contribution to the basin energies is also different for atomic basins containing various isotopes of an atom.[13-18] On the other hand, and from a computational viewpoint, the use of a single s-type Gaussian function as a nuclear basis set is common in ab initio calculations,[13-18] and when employed in conjunction with the Hartree type nuclear wavefunction,[31] equation (19) could used directly without any computational MC-QTAIM analysis to derive the contribution of nuclear vibrational energies to the basin energies.

### C. The anisotropic anharmonic oscillator model of the quantum nuclei

The isotropic harmonic oscillator model though offers a unique opportunity to derive a simple analytical equation, is just a simplified representation of complicated nuclear vibrations. Very recently, it has been demonstrated that the anisotropic anharmonic oscillator model maybe employed instead and within the context of this model a linear combination of various types of the Gaussian functions is used to represent $\psi_n(\vec{r}_n)$.[34] The nuclear basis set used in the original ab initio study employing the Hartree product wavefunction was a [2s2p2d] Gaussian basis set,[31] designed especially for the hydrogen isotopes,[19] which is the result of a specific anisotropic anharmonic oscillator model for $V_n^{ext}$.[34] The general structure of this basis set is used in this subsection as an example of employing the anisotropic anharmonic oscillator model to derive $\psi_n(\vec{r}_n)$.

In order to have sufficient computational flexibility and encompass the previously used nuclear basis sets in ab initio calculations,[19] a general basis set containing an arbitrary number of the s-, p- and d-type Cartesian Gaussian functions are used to expand $\psi_n(\vec{r}_n)$:

$$\psi_n(\vec{r}_n) = \sum_i^{N_s} a_i g_{s,i} + \sum_k^{N_p} \left( b_k g_{p_x,k} + c_k g_{p_y,k} + d_k g_{p_z,k} \right)$$
$$+ \sum_q^{N_d} \left( e_q g_{d_{xx},q} + f_q g_{d_{yy},q} + h_q g_{d_{zz},q} + r_q g_{d_{xy},q} + s_q g_{d_{xz},q} + t_q g_{d_{yz},q} \right)$$

$$g_{s,i} = \left(\frac{8\alpha_i^3}{\pi^3}\right)^{\frac{1}{4}} \exp\left(-\alpha_i \left|\vec{r}_n - \vec{R}_{n,c}\right|^2\right)$$

$$g_{p_{x,k}} = \left(\frac{128\beta_k^5}{\pi^3}\right)^{\frac{1}{4}} x \exp\left(-\beta_k \left|\vec{r}_n - \vec{R}_{n,c}\right|^2\right), \quad g_{p_y}^k \text{ and } g_{p_z}^k \text{ similarly}$$

$$g_{d_{xx},q} = \left(\frac{2048\gamma_q^7}{9\pi^3}\right)^{\frac{1}{4}} x^2 \exp\left(-\gamma_q \left|\vec{r}_n - \vec{R}_{n,c}\right|^2\right), \quad g_{d_{yy}}^q \text{ and } g_{d_{zz}}^q \text{ similarly}$$

$$g_{d_{xy},q} = \left(\frac{2048\gamma_q^7}{\pi^3}\right)^{\frac{1}{4}} xy \exp\left(-\gamma_q \left|\vec{r}_n - \vec{R}_{n,c}\right|^2\right), \quad g_{d_{xz}}^q \text{ and } g_{d_{yz}}^q \text{ similarly} \quad (20)$$

In these Gaussian functions, the subscript $n$ has been dropped for brevity and the enumeration of the functions is done using a new superscript. Employing equations (16), (17) and (20), and after some mathematical manipulations, one arrives at the nuclear contribution to the basin energies:

$$E(\Omega) = -\left(\frac{2\sqrt{2}\hbar^2}{m}\right)\left[\sum_i^{N_s}\sum_j^{N_s} a_i a_j \frac{3(\alpha_i \alpha_j)^{\frac{7}{4}}}{(\alpha_i + \alpha_j)^{\frac{5}{2}}} + \sum_k^{N_p}\sum_l^{N_p} (b_k b_l + c_k c_l + d_k d_l)\frac{10(\beta_i \beta_j)^{\frac{9}{4}}}{(\beta_i + \beta_j)^{\frac{7}{2}}}\right.$$

$$+ \sum_i^{N_s}\sum_q^{N_d} a_i (e_q + f_q + h_q)\frac{4(\alpha_i \gamma_q)^{\frac{7}{4}}(3\gamma_q - 2\alpha_i)}{\sqrt{3}(\alpha_i + \gamma_q)^{\frac{7}{2}}}$$

$$\left. + \sum_q^{N_d}\sum_p^{N_d} (e_p e_q + f_p f_q + h_p h_q)\frac{4(\gamma_p \gamma_q)^{\frac{7}{4}}(17\gamma_p \gamma_q - 2\gamma_p^2 - 2\gamma_q^2)}{3(\gamma_p + \gamma_q)^{\frac{9}{2}}}\right.$$

$$+\sum_{q}^{N_d}\sum_{p}^{N_d}\left(e_p f_q + e_p h_q + f_p h_q\right)\frac{8\left(\gamma_p\gamma_q\right)^{\frac{7}{4}}\left(3\gamma_p\gamma_q - 2\gamma_p^2 - 2\gamma_q^2\right)}{3\left(\gamma_p + \gamma_q\right)^{\frac{9}{2}}}$$

$$\left.+\sum_{q}^{N_d}\sum_{p}^{N_d}\left(r_p r_q + s_p s_q + t_p t_q\right)\frac{28\left(\gamma_p\gamma_q\right)^{\frac{11}{4}}}{\left(\gamma_p + \gamma_q\right)^{\frac{9}{2}}}\right] \quad (21)$$

Clearly, this equation is more complicated than equation (19) but it has the required flexibility to include subtle effects emerging from the anisotropy and the anharmonicity of the nuclear vibrations. If a specific basis set is used in an ab initio calculation and the optimized linear coefficients and the exponents of the basis set are determined in a variational calculation, then equation (21) could be used to derive the contribution of the nuclear vibrations to the basin energy directly without any computational MC-QTAIM analysis.

**D. A "heuristic" approach toward inclusion of the zero-point vibrational energies into the QTAIM**

The developed formalism in the previous subsections and particularly equation (16) set the stage to evaluate the contribution of the zero-point nuclear vibrational energy to the basin energy analytically where equations (19) and (21) were vivid examples. The contribution of the nuclear vibrational energies to the basin energies was one of the missing ingredients of the QTAIM and now, within the context of the MC-QTAIM, this is understandable; the orthodox formalism is based on the electronic Hamiltonian where the clamped nucleus model is assumed from the outset.[1] In other words, the MC-QTAIM is the "natural" framework to incorporate the nuclear contributions into various basin properties since it is based on a more general Hamiltonian, equation (1), where certain nuclei are treated as quantum waves. However, inspired by the MC-QTAIM formalism, it is interesting to contemplate how one may "heuristically" incorporate the vibrational energies into the orthodox formalism.

In the adiabatic view, the molecular Hamiltonian is divided to the electronic and the nuclear Hamiltonians therefore the molecular wavefunction is the product of the electronic and the nuclear wavefunctions. In this framework, the electronic Schrodinger equation is solved at first step, $\hat{H}_e \Psi_e = E_e \Psi_e$, and then its eigenvalues are employed as the potential energy surfaces for the nuclear Schrödinger equation. The nuclear Schrödinger equation may itself be divided to translational, rotational, and vibrational Schrödinger equations, $\hat{H}_{vib} \Psi_{vib} = E_{vib} \Psi_{vib}$, if the coupling between rotational and vibrational motions is negligible. The electronic wavefunction is used as input to the QTAIM analysis thus the resulting boundaries of the atomic basins and their properties are completely unaffected by the vibrational wavefunction. With this background in mind, it seems legitimate to suppose that the basin contributions of the vibrational energies of nuclei must be added to the "predetermined" electronic basin energies derived from the formalism of the QTAIM at the equilibrium geometry,[1] to yield the total basin energies: $E_t(\Omega) = E_e(\Omega) + E_{vib}(\Omega)$, $E_e = \sum_\Omega E_e(\Omega)$, $E_{vib} = \sum_\Omega E_{vib}(\Omega)$. Accordingly, extrapolating the reasoning of subsection II.B, a *vibrational energy density* must be introduced that upon basin integration, yields the basin contribution of the vibrational energies: $E_{vib}(\Omega) = \int_\Omega d\vec{q}\, E_{vib}(\vec{q})$.

In general we have no recipe from the orthodox formalism to construct the proper vibrational energy density.[1] However, it has been demonstrated very recently that the partitioning based on the virial projection operator is extendable to the whole class of the homogenous potential energy functions, $\hat{V}(s\vec{q}_1,...,s\vec{q}_N) = s^k \hat{V}(\vec{q}_1,...,\vec{q}_N)$, where the coulombic potential energy function is just an example, $k = -1$.[35] On the other hand, the usual vibrational Hamiltonian is based on the model of coupled harmonic oscillators,[36] which also belongs to this class, $k = 2$.[35] Accordingly, introducing vibrational energy density based on the same formalism which was used to introduce the energy density for the coulombic potentials seems feasible. In this paper, as a proof of concept, only a "toy model" for vibrations based on a set of uncoupled harmonic oscillators is considered in detail, neglecting the translational and rotational contributions, leaving a comprehensive analysis of the more reliable model of coupled harmonic oscillators to a future report.

Let's introduce the following vibrational Hamiltonian assuming there are $m$ distinguishable nuclei and each nucleus experiences a harmonic potential energy surface locally:

$$\hat{H}_{vib} = \sum_i^m \left(\frac{-\hbar^2}{2m_i}\right)\nabla_i^2 + \hat{V}, \quad \hat{V}(\vec{q}_1,...,\vec{q}_N) = \frac{1}{2}\sum_i^m k_i |\vec{q}_i|^2 \qquad (22)$$

In this Hamiltonian $\vec{q}_i = \vec{r}_i - \vec{r}_{i,e}$ where $\vec{r}_{i,e}$ is the equilibrium position of the $i-th$ nucleus and $k_i$ is the corresponding force constant. The potential energy surface of the system is a homogenous function and the corresponding virial projection operator is as follows: $\hat{V} = \left(\frac{1}{2}\right)\sum_i^m \vec{q}_i \cdot \vec{\nabla}_i \hat{V}$.[35] Using this projection operator, the vibrational Hamiltonian transformed as the following:

$$\hat{H}_{vib} = \sum_i^m \hat{h}_i = \sum_i^m \left\{\left(\frac{-\hbar^2}{2m_i}\right)\nabla_i^2 + \left(\frac{1}{2}\right)\vec{q}_i \cdot \vec{\nabla}_i \hat{V}\right\} \qquad (23)$$

This equation set the stage, similar in sprit to equation (18), to introduce the proper energy density, similar in spirit to equation (11), as the following:

$$E_{vib}(\vec{q}) = T_{vib}(\vec{q}) - \left(\frac{1}{2}\right)V_{vib}^T(\vec{q})$$

$$V_{vib}^T(\vec{q}) = V_{vib}^B(\vec{q}) + V_{vib}^S(\vec{q}), \quad T_{vib}(\vec{q}) = \sum_i^m T_{vib}^i(\vec{q}), \quad V_{vib}^B(\vec{q}) = \sum_i^m V_{vib,i}^B(\vec{q})$$

$$V_{vib}^S(\vec{q}) = \sum_i^m V_{vib,i}^S(\vec{q}), \quad T_{vib}^i(\vec{q}_i) = \int d\tau_i' \, \Psi_{vib}^*\left(-\frac{\hbar^2}{2m_i}\right)\nabla_i^2 \Psi_{vib},$$

$$V_{vib,i}^B(\vec{q}_i) = \int d\tau_i' \, \Psi^*\left(-\vec{q}_i \cdot \vec{\nabla}_i \hat{V}\right)\Psi, \quad V_{vib,i}^S(\vec{q}_i) = \vec{\nabla}_i \cdot \left(\vec{q}_i \bullet \ddot{\sigma}_{vib,i}(\vec{q}_i)\right) \quad (24)$$

In these equations $d\tau_i'$ implies integrating over spatial coordinates of all nuclear variables except the coordinate of $i-th$ nucleus. Also, $\ddot{\sigma}_{vib,i}(\vec{q}_i)$ is the Schrodinger-Pauli-Epstein stress tensor density of $i-th$ nucleus defined as follows:

$$\ddot{\sigma}_{vib,i}(\vec{q}_i) = \left(\frac{\hbar^2}{4m_i}\right)\int d\tau_i' \left\{\Psi_{vib}^*\left(\vec{\nabla}\bullet\vec{\nabla}\Psi_{vib}\right) + \Psi_{vib}\left(\vec{\nabla}\bullet\vec{\nabla}\Psi_{vib}^*\right) - \left(\vec{\nabla}\Psi_{vib}^*\right)\bullet\left(\vec{\nabla}\Psi_{vib}\right) - \left(\vec{\nabla}\Psi_{vib}\right)\bullet\left(\vec{\nabla}\Psi_{vib}^*\right)\right\}.$$

At this stage of development, the formalism is complete and using the vibrational wavefunctions derived from the Schrödinger equation the energy density must be calculated explicitly and then integrated within the atomic basins delineated through the QTAIM analysis. However, for the case of the localized vibrations, the formalism maybe simplified further using the local virial theorem of $k = 2$ systems.

In some previous studies, [35,37,38] it was demonstrated that the local virial theorem is not tied to the coulombic interactions and can be used for the whole class of the systems with the homogenous potential energy functions. Thus, similar to equations (14), the local virial theorem for the system under study is as the following:

$$2T_{vib}^{i}(\vec{q}_i) = -V_{vib,i}^{T}(\vec{q}_i) - \left(\frac{\hbar^2}{4m_i}\right)\nabla^2 \rho_i(\vec{q}_i), \quad i=1,...m$$

$$V_{vib,i}^{T}(\vec{q}_i) = V_{vib,i}^{B}(\vec{q}_i) + V_{vib,i}^{S}(\vec{q}_i), \qquad \rho_i(\vec{q}_i) = \int d\tau_i' \, \Psi_{vib}^{*} \Psi_{vib} \qquad (25)$$

Since each nucleus resides in a single atomic basin, $\Omega_i$, and throughout this paper the localized nature of nuclear distribution in the class A systems is assumed, then: $\int_{\Omega_i} d\vec{q}_i \, \nabla^2 \rho_i(\vec{q}_i) \approx 0$, $\vec{\sigma}_{vib,i}(\partial\Omega_i) \approx 0$, and so: $V_{vib,i}^{S}(\Omega_n) = \int_{\Omega_i} d\vec{q}_i \, \vec{\nabla}_i \cdot (\vec{q}_i \bullet \vec{\sigma}_{vib,i}(\vec{q}_i)) = \oint_{\partial\Omega_i} dS (\vec{q}_i \bullet \vec{\sigma}_{vib,i}(\vec{q}_i)).\vec{n} \approx 0$.

Accordingly, combining equation (24) and (25) and integrating within the atomic basins the following equation is derived for the localized vibrations:

$$E_{vib}(\Omega_i) = 2T_{vib}(\Omega_i) = -V_{vib}^{B}(\Omega_i)$$

$$T_{vib}(\Omega_i) = \int_{\Omega_i} d\vec{q} \, T_{vib}(\vec{q}), \qquad V_{vib}^{B}(\Omega_i) = \int_{\Omega_i} d\vec{q} \, V_{vib}^{B}(\vec{q}) \qquad (26)$$

This equation offers a simple way to incorporate zero-point vibrational energies into the atomic basin energies just knowing the ground state kinetic or basin virial energy densities of the vibrating nuclei. In order to evaluate the energy density for the model of uncoupled harmonic oscillators, the ground state wavefunction of the model, $\Psi_{vib} = \prod_{i}^{m} \phi_{vib,i}(\vec{q}_i)$,

$\phi_{vib,i}(\vec{q}_i) = \left(\frac{\sqrt{m_i k_i}}{\hbar \pi}\right)^{\frac{3}{4}} \exp\left(-\frac{\sqrt{m_i k_i}}{2\hbar}|\vec{q}_i|^2\right)$ is used to compute the basin contribution of the vibrational zero-point energies. After some mathematical manipulations, one arrives at:

$$E_{vib}(\Omega_i) = \left(\frac{3\hbar}{2}\right)\sqrt{\frac{k_i}{m_i}} \qquad (27)$$

This is the zero-point energy of a single 3D isotropic harmonic oscillator revealing the expected "local" nature of vibrations namely, each nucleus is a quantum oscillator confined to its own atomic basin; the total basin energy is the sum of the electronic basin energy emerging from the QTAIM analysis and the vibrational energy of the nucleus resides in the basin, which seems to be an appealing physical picture. This localized nature is even more evident if the vibrational wavefunctions of any excited state of the introduced model are used to compute the regional vibrational energy; in such cases, the extra energy of the vibrational excitation contributes only to those basins containing the vibrationally excited nuclei. In a more advanced study on the nuclear vibrations, the normal mode analysis must be introduced within the context of the coupled harmonic oscillator model,[39] and the vibrational energy contribution of a nucleus from each normal mode must be disentangled and added together in order to reach a formalism similar to the present toy model (Shahbazian, under preparation).

## IV. Conclusion and Prospects

To be fair, the effect of nuclear vibrations on the QTAIM analysis had been appreciated by Bader and coworkers since the start of developing the theory.[1] However, the role of nuclear vibrations was confined to the topological analysis of the electron density; since the electronic wavefunctions and corresponding electron densities depend on nuclear coordinates parametrically, the gradient vector field of the electron density is distinct for each molecular geometry.[1] In other words, the topological analysis could be done not only at the equilibrium geometry, but also at non-equilibrium geometries and this opens the door to pursue the evolution of the topology of the electron density, e.g. molecular graphs, because of the zero-point nuclear vibrations,[40-43] as well as during the course of chemical reactions.[1] However as also emphasized previously, beyond the topological analysis, the effect of nuclear vibrations on the basin properties had not been considered within the context of the QTAIM analysis and this is even more pronounced for the basin energies since the electronic basin energies are only defined at the equilibrium geometries. The present study demonstrates that if one is interested to incorporate the nuclear vibrational contribution into the basin energies then the formalism must be inevitably extended; the MC-QTAIM formalism naturally includes nuclear vibrations not only at the level of the topological analysis, through introducing the Gamma density, but also at the level of basin properties, through introducing combined property densities. However, a heuristic method, like that proposed in this paper, may also be used to estimate the contribution of the vibrational energies within the context of the QTAIM though proposing such a method is also an extension of the conventional formalism through introducing the vibrational energy density as a new ingredient. Let's have a view to future possibilities that these extensions are promising.

It was intellectually appealing to incorporate nuclear vibrational energies into the AIM analysis, however, beyond academic curiosity, this achievement opens the door for a direct comparison of the predictions of the QTAIM on the contribution of the functional groups to thermodynamic functions and corresponding experimentally derived group additive values.[1] The tradition of decomposing a molecular property, e.g. energy or polarizability, into contributions of the functional groups constituting the molecule has a long history in organic chemistry, which has been well documented in the pioneering studies of Benson and coworkers.[44-46] While Bader and coworkers have considered the idea of group additivities within the context of the QTAIM in detail,[1,47-49] any quantitative comparison to experimentally derived values needs to take the vibrational contribution to thermodynamic functions into account. The fact that any derivation of the group additive contributions from the usual ab initio quantum mechanical has an indirect nature,[50] the capability of deriving computational group additive values from the QTAIM analysis plus the proposed heuristic vibrational correction is rewarding. The modern literature on the additivity schemes is quite rich,[51-70] and the prospect for a detailed comparison to experimentally derived values seems to be promising. This line of research will be pursued computationally after introducing the model of coupled harmonic oscillators in a future report.

The Hartree product nuclear wavefunction used in present study though yield proper results within the context of the MC-QTAIM analysis, lacks the electron-nucleus correlation which is an important ingredient if one tries to derive the vibrational frequencies quantitatively from non-adiabatic ab initio calculations. Thus, in next step it is desirable to design simple (minimal) wavefunctions that capture the main "physics" of this correlation through introducing the explicit correlation of the quantum nuclei and electrons into the wavefunction with a proper correlation factor.[33,71-79] The linear combination of the explicitly correlated Gaussian functions used by Hammes-Schiffer and coworkers as the correlation factor is a natural choice for future analytical studies though other promising mathematical functions are also worth trying. In the ideal situation one may hope to find a wavefunction simple enough to be used for the analytical studies and flexible enough to be able to reproduce the correct values of vibrational frequencies in computational studies. This is a challenging line of research that is now under investigation in our laboratory.

The next step for the incorporation of nuclear vibrational energies into the QTAIM analysis, after considering the model of the coupled harmonic oscillators, is going beyond the harmonic oscillator model and try to introduce the energetic contribution of anharmonicity regionally, though, this is probably an order of magnitude smaller than the harmonic contribution. Since the anharmonic models of vibrations in general are not the members of the class of the homogenous potential energy functions,[36,39] the strategy used in this paper is not directly applicable to these types of models. Accordingly, a more general strategy to construct energy densities needs to be introduced, not tied to the homogeneity of the used potential energy function. Such possibility will be considered in a future study.

Finally, one may conclude that while analytical studies are scare in the domain of the QTAIM, the present and previous analytical studies within the context of the MC-QTAIM demonstrate that analytical studies reveal aspects of the AIM analysis, which are hard to be grasped solely by the computational studies. A fresh look at the QTAIM formalism, through constructing analytical models, seems to be a promising path to be pursued in future studies.

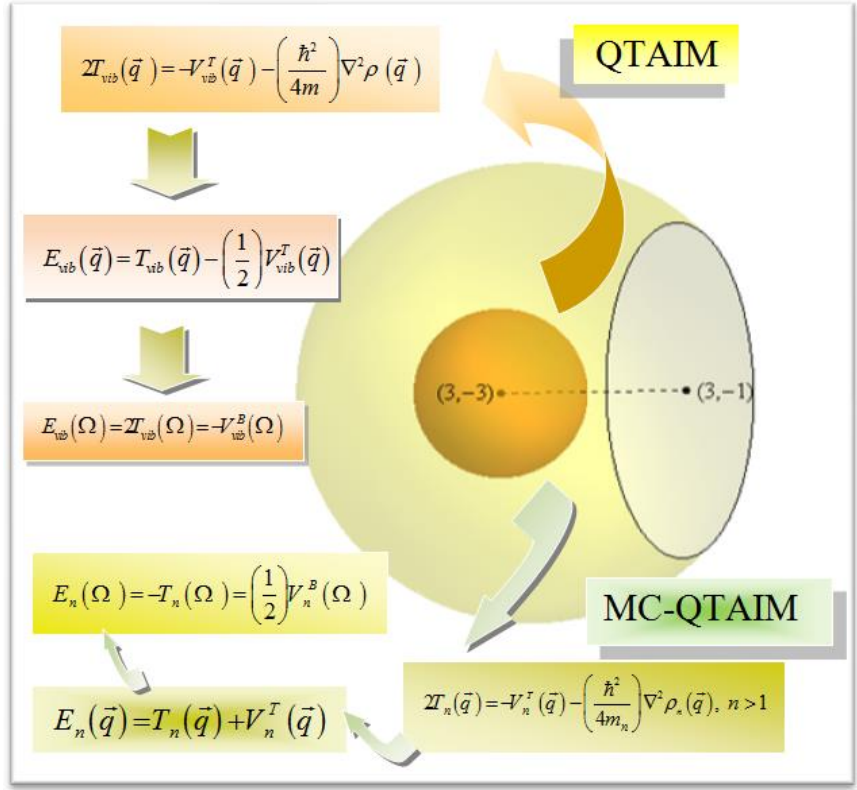